\documentclass[epj-ND]{svjour}
\usepackage[tbtags]{amsmath}
\usepackage{amssymb}
\usepackage{graphicx}

\usepackage{txfonts}

\def\fpago{$\mathsf{^{19}F(p,\alpha\gamma)^{16}O}$}

\begin{document}

\title{\boldmath{\fpago} reaction: HPGe detector response function and
  gamma-ray relative yields}

\author{
  D.~B.~Tridapalli \inst{1} \fnmsep \thanks{\email{dtridapa@if.usp.br}} \and
  P.~R.~Pascholati \inst{1} \fnmsep \thanks{Presenting author} \and
  N.~L.~Maidana    \inst{1} \and
  V.~R.~Vanin      \inst{1} \and
  Z.~O.~Guimar\~aes-Filho \inst{1} \and
  M.~A.~Rizzutto \inst{2} 
}

\institute{
  Laborat\'orio do Acelerador Linear, Universidade de S\~ao
  Paulo, S\~ao Paulo, Brasil
  \and
  Laborat\'orio An\'alise de Materiais  por Feixes I\^onicos, Universidade de 
  S\~ao Paulo, S\~ao Paulo, Brasil
}


\abstract{
  The 6.1, 6.9 and  7.1~MeV photon yields in the reaction with protons in the
  1.32 to 1.42~MeV energy range were determined from the comparison between
  experimental and simulated spectrum. The gamma-ray spectra were measured with
  a HPGe detector which was 72.5~mm in diameter and 60.5~mm in length. The
  reaction kinematics and the detector response function were simulated by
  different programs using the Monte Carlo method. The relative photon yields
  were obtained from the least-squares fit of the experimental gamma-ray
  spectra to the convolution of the detector response function with the photon
  energy distribution arising from the reaction kinematics.  
}

\maketitle

\section{Introduction}

The \fpago\ reaction is used in many areas, from
detector~\cite{guldbakke:90,croft:91a} and dosimeter~\cite{duvall:88,rogers:83}
calibration to non destructive analysis of fluorine trace with
PIGE~\cite{jarjis:78,dieumengard:80}. It was noted by many
authors~\cite{spyrou:97,fessler:00,dababneh:93,croft:91b,ding:02,ajzenberg-selove:77}
that the available data as reaction cross section, photon yields, resonance
energies are insufficient and in some cases inconsistent.

The HPGe response function for photons presents a strong dependence on energy
due to the different dominant interaction processes according to the energy
region. In some studies, the photon detector response is described by means of
empirical functions~\cite{jin:86,lee:87}, but this method cannot be extended to
energies above those used to fit the empirical parameters. Monte Carlo
simulation, owing to the increasing computer power, is a method in growing
adoption~\cite{buermann:93}, because of the detailed accounting of the physical
interaction processes. We use \emph{Monte Carlo N-Particle}, version 5
(MCNP5)~\cite{mcnp} to calculate the HPGe photon detector response function.

Nuclear structure properties along with the reaction mechanism and kinematics
lead to Doppler broadening and shift of the emitted photons. A computer program
was developed to simulate the gamma-ray line profile. The convolution of the
simulated profile with the detector response function was fitted to the
experimental spectra; from this fit, the high-energy gamma-ray yields in the
\fpago\ reaction could be deduced, and the adequacy of the detector response
function was tested.

\section{Experimental method}

The experiment was performed with the 1.7~MV Pelletron tandem accelerator of
the Laborat\'orio de An\'alise de Materiais por Feixes I\^onicos
(LAMFI)~\cite{LAMFI}  located at Instituto de F\'isica da Universidade de S\~ao
Paulo, Brasil.  

\begin{figure}[b]
  \begin{center}
    \resizebox{0.95\columnwidth}{!}{
      \includegraphics{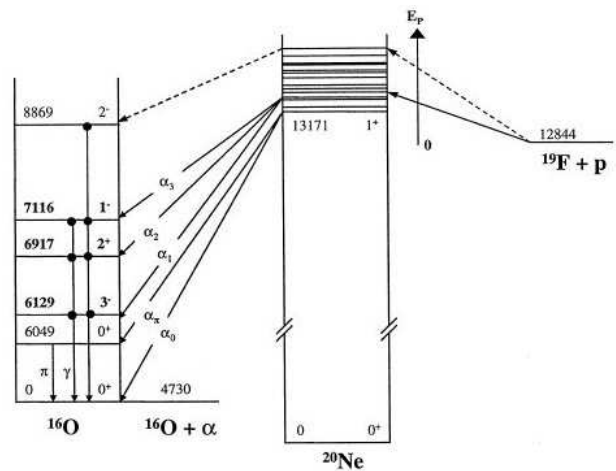}
    }
    \caption{The reaction mechanism, showing simplified level schemes of
      $^{16}\textrm{O}$ and $^{20}\textrm{Ne}$. All the energies are in keV;
      level schemes not drawn to scale.}
    \label{fig:levels}
  \end{center}
\end{figure}

The target, consisting of about 260~$\mu$g/cm$^{2}$ CaF$_{2}$ evaporated on a
0.1~mm Ta backing, was irradiated with protons in the 1.362~MeV to 1.416~MeV
energy range. The proton energy loss in the target thickness is about 37~keV;
hence, the proton resonances at 1.34~MeV and~1.37 MeV, with widths of 4.8(8)
and 11.9(12)~keV respectively, were excited~\cite{spyrou:97}. Gamma-rays with
energies: 6.1~MeV, 6.9~MeV and 7.1~MeV from transitions in the  reaction
product were observed; fig.~\ref{fig:levels} depicts the reaction
mechanism.

The detector system consisted in a reverse-electrode closed-end coaxial HPGe
detector, CANBERRA model GR6024, measuring 72.5~mm in diameter and 60.5~mm in
length, and conventional electronics. The detector was placed with its axis
along the beam direction, with the detector capsule at 12.2~cm from the target.

Thirteen measurements were done with incident proton energies varying from
1.362~MeV to 1.416~MeV, where the live counting time for each spectrum was
600~s. The proton irradiation current was around 50~nA and the beam size was
about 3~mm. 

\section{Simulation}
In order to calculate the observed photon spectrum in the HPGe detector, two
different simulations were performed. First, the detailed reaction process was
simulated to account for the Doppler broadening and shift of the photons
emitted from the $^{16}$O. Then, a beam of photons with the energy distribution
obtained in the first simulation was used in the MCNP5 source input card to
determine the HPGE detector response function. 

\subsection{Reaction mechanism and emitted photon spectrum}
In the \fpago\ reaction the gammas could be emitted in flight. In order to
determine the direction and velocity of the $^{16}$O nuclei at the emission
moment was made a Monte Carlo simulation of the reaction
kinematics~\cite{diogo}.

The $^{20}$Ne resonant state fissions, with very short half life
($\approx10^{-20}$s), populating different $^{16}$O excited states. The level
energies, half-lives and gamma-ray transitions of the $^{16}$O excited states
observed in the experiment are given in table~\ref{tab:half-life} and, although
much grater than $^{20}$Ne lifetime, are shorter than the time required for
stopping the residual nuclei in the target. Hence, the observed gamma-ray peaks
are Doppler shifted and broadened due to the $^{16}$O nucleus motion at the
emission time.

\begin{table}[t]
  \begin{center}
    \caption{$^{16}$O level energies ($E_r$), half lives and gamma transition
      energies ($E_\gamma$)~\cite{nndc:nudat}.}
    \label{tab:half-life}
    \begin{tabular}{lll}
      \hline\noalign{\smallskip}
      $E_r$ (keV) & Half life & $E_\gamma$ (keV) \\
      \noalign{\smallskip}\hline\noalign{\smallskip}
      6 129,89 (4) & 18,4 (5)~ps &  6 128,63 (4) \\
      6 917,1 (6) & 4,70 (13)~fs &  6 915,5 (6) \\
      7 116,85 (14) & 8,3 (5)~fs &  7 115,15 (14) \\
      \noalign{\smallskip}\hline
    \end{tabular}
  \end{center}
\end{table}

The reaction simulation starts by the random sorting of the target point where
the interaction occurs, considering the reaction cross-section and the proton
energy loss along its way through the target. In this point, $^{20}$Ne
formation and prompt fission occurs; the resulting $^{16}$O oxygen nucleus,
emitted in isotropically sorted direction in the center of mass reference
frame, begins its trajectory along the CaF$_2$ and tantalum  target layers or
into the irradiation chamber inner void. A different run is performed for each
excited state, in order to follow the $^{16}$O kinetic energy loss and multiple
scattering through its trajectory; this path is followed for a time interval
whose length is sorted from the corresponding decay time distribution. Finally,
the gamma-ray is emitted, assuming isotropic angular distribution, and the
energy spectrum is obtained.

\subsection{Detector response function}
\label{sec:response}
The HPGe detector response function was simulated using the tally F8:P (pulse
height) of MCNP5, and a small (1~keV) channel width to compare this spectrum
with the experimental one. A different photon source was used for each $^{16}$O
gamma-ray transition, corresponding to the photon energy distributions
calculated by the reaction simulation, as explained in the preceding
sub-section. Therefore, the MCNP output tally is already the convolution of the
detector response function with the reaction gamma-ray line profile, Doppler 
broadened and shifted.

The detector model included, besides the active Ge volume with its inner hole
and the germanium crystal dead layers, many detector assembly elements: the
beryllium window, aluminium internal and external capsules, the dewar filled
with liquid nitrogen and the cold finger. Some details with low mass or placed
away the incident radiation, like the pre-amplifier electronic circuits, were
not considered. To achieve a good accuracy of the position and thicknesses of
detector inner components, a radiography of the detector was taken. It was
found that the detector symmetry axis was tilted relative to the detector
capsule~\cite{diogo}, and the detector model was changed accordingly.

A simplified design of the irradiation chamber and its metallic support, with a
detailed design of the target holder, were considered in the
simulation. Scattered photons were followed inside a 3~m sphere centered in the
target, covered by 5~cm concrete walls at all directions but the ground floor,
which cut the bottom part of this sphere at 67~centimeters below the beam line.

The photon angular distribution was considered isotropic, but the small solid
angle covered by the detector, about 0.3~sr, made a full isotropic simulation
prohibitively time consuming. Therefore, the simulation was made in two steps: 

\begin{enumerate}
  \item Spectrum simulation with the photons emitted with uniform angular
    distribution within the detector solid angle.
  \item Spectrum simulation with the photons emitted with uniform angular
    distribution in all directions outside those covered by the detector.
\end{enumerate}

The 6.9~MeV and 7.1~MeV reaction gamma-rays calculated spectra did not change
appreciably in the range of proton irradiation energies used in this
experiment,  because their short half-lives imply small paths in the target;
therefore, the same simulated spectra were used in the comparison with the
experimental data of all thirteen irradiations. On the contrary, the 6.1~MeV
gamma-rays are emitted from a nuclear state with greater half-life, requiring
detailed account of the target point where the interaction occurs, because the
$^{16}$O nucleus can cross the boundaries of the CaF$_2$ target layer either to
the Ta backing or to the scattering chamber void. Since the distribution of the
interaction point in the target changes with the proton incident energy, a
spectrum for each one of the 13 different irradiation proton energies was
calculated, for gamma-rays emitted  inside the detector solid angle. For
gamma-rays emitted outside the solid angle covered by the detector, whose
contribution to the observed spectrum is secondary, a single simulation was
performed. 

\section{Comparison between experiment and simulation}
The electronics detector system contribution to the experimental gamma-ray
spectrum resolution must be taken into account in the comparison
with the experiment. The usual model for the dependence of the peak Full Width
at Half Maximum (FWHM) with gamma-ray energy~\cite{knoll} could not be applied
to this case because the very prominent escape peaks do not follow this
model. We adopted a simplified model, assuming that all the peaks have
the same FWHM, which becomes a fitting parameter. Therefore, the electronics
noise function was assumed gaussian with the same FWHM for all energies.

The model used to compare the simulated and experimental spectra is a
linear combination of the convolution of the simulated response functions of
the preceding section with the electronics noise function plus a constant
background: 

\begin{eqnarray}
  C(E) = I_{6.1}\cdot S_{6.1}(E,W) +
    I_{6.9}\cdot S_{6.9}(E,W) + \nonumber \\
    + I_{7.1}\cdot S_{7.1}(E,W) + B
  \label{eq:lsf}
\end{eqnarray}

\noindent where $C(E)$ is the number of counts in the channel corresponding to
energy $E$, $I_{6.1}$, $I_{6.9}$ and $I_{7.1}$ are the gamma-ray intensity
parameters, $B$, a constant background, and $S_i$, the convolutions of the
simulated response functions for the gamma-rays with
energy $i$ with the electronics noise function, given by:

\begin{equation}
  S_{i} (E,W) = f_i(E,W) + \Omega \cdot fc_i(E,W)
\end{equation}

\noindent where $f_i(E,W)$ and $fc_i(E,W)$ correspond to the gamma-rays emitted
inside and outside the solid angle covered by the detector, respectively, and
$\Omega=40$ is the fixed ratio between the solid angle covered by detector and
its complement to 4$\pi$.

\begin{figure}
  \begin{center}
    \resizebox{0.95\columnwidth}{!}{
      \includegraphics{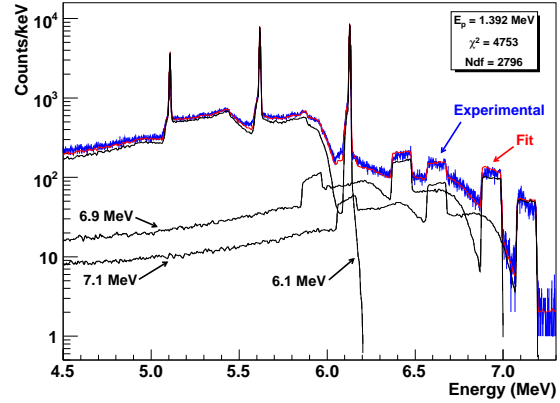}
    }
    \caption{Calculated and experimental spectra obtained for  
    $E_p$ = 1.392~MeV. The components $I_{i}\cdot S_{i}(E,W)$ of 
    eq.~\eqref{eq:lsf} are also drawn separately.}
    \label{fig:spectrum}
  \end{center}
\end{figure}

A set of $I_{6.1}$, $I_{6.9}$, $I_{7.1}$, $W$ and $B$ parameters was fitted for
each incident proton energy. The fitted gamma-ray spectrum energy ranges
from 4.5~MeV to 7.3~MeV, which is a larger energy region than those analysed by other
authors~\cite{duvall:88,fessler:00,micklich:03}, which fit the photon spectrum
above 5~MeV. The calculated and experimental spectra obtained with protons of
1.39~MeV can be seen in Fig.~\ref{fig:spectrum}.

The obtained reduced chi-squares, $\chi^2_R$, are between 1.3 and 1.9, outside
the acceptance region from a statistical point of view. These high $\chi^2_R$
may be the result of: the underestimation of the continuum part of the photon
spectrum by MCNP5 simulations, which was also noted in other
works~\cite{utsunomiya:05}; the description of the peaks, with the
simplified modeling of the elecronics noise; and the 
inaccurate reaction kinematic parameter values, like target
thickness and incident proton energy.

\section{Discussion}

The yields of the reaction gamma-rays were obtained by other authors
refs.~\cite{duvall:88,fessler:00,micklich:03} with similar procedures. However,
the scintillation detectors employed, with poor energy resolution when compared
to HPGe detectors, did not allow the observation of the peak broadening due to
photon emission in flight.

The underestimation of the continuum part of the gamma-ray spectrum was
assigned by other authors to the limitations of the experimental arrangement 
geometric design included in the simulation  ~\cite{duvall:88,fessler:00}, 
composed of just the source and the detector. In this work, the experimental
arrangement was considered in detail, see sec.~\ref{sec:response}, therefore
that explanation is not applicable to our case. Indeed, the detailed modeling
of the detector arrangement allowed the increase in the experimental gamma-ray
spectrum fitting region, notwithstanding the systematic overestimation of the
continuum part of the experimental spectrum. Therefore, our results indicate
that the simulation programs underestimate the continuum, as previously pointed
by Utsunomiya et. al.~\cite{utsunomiya:05}, who assigns this phenomenon to
crystal Ge surface inhomogeneities, not taken into account in the simulations,
that displace events from the full-energy peak to the continuum. 

Several tests were performed to check the stability of the results against 
changes in the fitted gamma-ray spectrum energy range and peak FWHM. The fit
was also repeated with the 6.1~MeV peak and corresponding annihilation escape
peaks removed from the fitting procedure. In all cases, the obtained parameters
were compatible, showing that our analysis methodology was
not sensible to these changes.

\begin{figure}
  \begin{center}
    \resizebox{0.95\columnwidth}{!}{
      \includegraphics{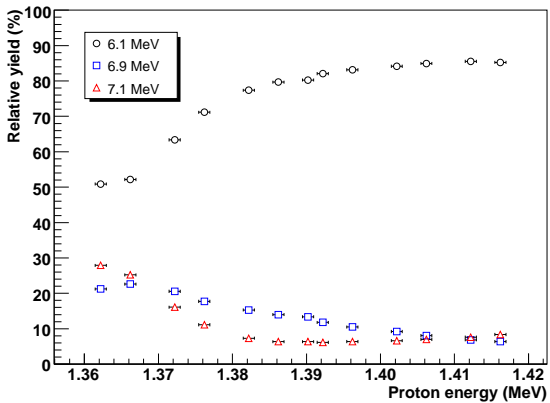}
    }
    \caption{\fpago\ reaction photon relative yield as function of the proton
      incident energy. The y-error bars are lower than the points.}
    \label{fig:relative_yield}
  \end{center}
\end{figure}

The relative gamma-ray yields, calculated by:

\begin{equation}
  Y_{r_i} = \frac{I_{i}}{I_{6.1}+I_{6.9}+I_{7.1}}
\end{equation}

\noindent where $I_i$ are the parameters in eq.~\eqref{eq:lsf} obtained from
the fit, are shown in fig.~\ref{fig:relative_yield}, in function of the proton
bombarding energy.

Although we could not measure absolute yields, the coulomb excitation
gamma-rays from the Ta backing provided a beam intensity reference, smoothly
varying with proton beam energy. Therefore, with the parameters $I_i$ from the
fit of the eq.~\eqref{eq:lsf} and the peak area of the 198~keV gamma-ray from
Ta coulomb excitation, $C_{Ta}$, the yield of the gamma-ray with energy $i$ is
given by the expression:

\begin{equation}
  Y_{i} = \frac{I_{i}}{C_{Ta}}.
\end{equation}

\begin{figure}
  \begin{center}
    \resizebox{0.95\columnwidth}{!}{
      \includegraphics{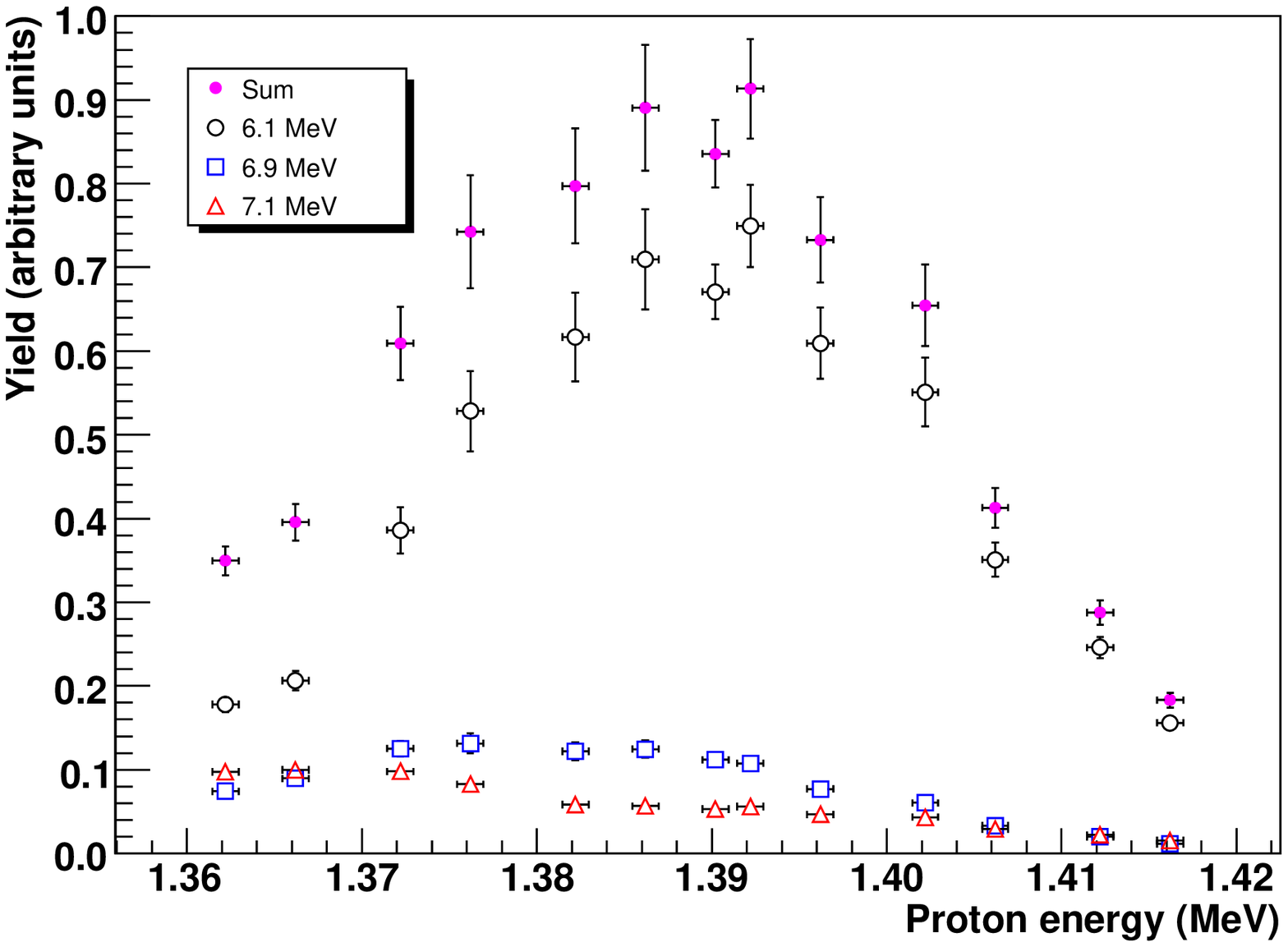}
    }
    \caption{\fpago\ reaction photon yield as function of proton incident
      energy.}
    \label{fig:yield}
  \end{center}
\end{figure}

The calculated gamma yields for the energies measured can be seen at
fig.~\ref{fig:yield}. The table containing the yield numerical values 
can be seen on page 59 of ref.~\cite{diogo}.

\section{Conclusion}
The data analysis methodology used in this work, although not new, was never
applied before to a high-resolution detector, like the HPGe detector used, 
to the best of our knowledge.

The \fpago\ reaction has proven to be useful in the HPGe detector calibration
at high energy, despite the Doppler broadening complications, which makes 
necessary significant additional work to model the reaction gamma-rays peak
shape, but allowed an accurate gamma-ray relative yield determination.

Our results indicate that the systematic underestimation of the continuum 
component of the experimental gamma-ray spectrum by simulation programs where
the photon and electron transport phenomena are dealed with in details,
cannot be always assigned to insufficient description of
the experimental arrangement; Utsunomiya et. al.~\cite{utsunomiya:05} assigns
this characteristic to a detection phenomenon still not taken into account by
the simulation programs.

\begin{acknowledgement}
We would like to thank the scientific and technical staff of Pelletron and
LAMFI Laboratories of USP for help in the irradiations; 
CNPq (Conselho Nacional de Desenvolvimento Cient\'ifico e Tecnol\'ogico,
Brazil) and FAPESP (Funda\c c\~ao de Amparo \`a Pesquisa do Estado de
S\~ao Paulo, Brazil) for partial financial support.
\end{acknowledgement}

\bibliographystyle{epj}
\bibliography{630_pascholati}

\end{document}